\documentclass[aps,prb,preprint]{revtex4}
\usepackage{graphicx}
\usepackage{amsmath}
\usepackage{scalerel,stackengine}
\stackMath
\newcommand\reallywidehat[1]{%
\savestack{\tmpbox}{\stretchto{%
  \scaleto{%
    \scalerel*[\widthof{\ensuremath{#1}}]{\kern-.6pt\bigwedge\kern-.6pt}%
    {\rule[-\textheight/2]{1ex}{\textheight}}
  }{\textheight}%
}{0.5ex}}%
\stackon[1pt]{#1}{\tmpbox}%
}

\begin{document}

\title{A centennial reappraisal of Heisenberg's Quantum Mechanics with a perspective on Einstein's Quantum Riddle.}

\author{Tuck C. Choy}
\email{tuckvk3cca@gmail.com}

\affiliation{Samy Maroun Center for Space, Time and the Quantum, Parc Maraveyre B${\hat a}$t. 1, 13260, Cassis, Bouches du Rh${\hat o}$ne, France  and \\
Departmento de Fisica-CIOyN, Universidad de Murcia, Murcia 30071, Spain.}


\begin{abstract}

Heisenberg's breakthrough in his July 1925 paper that set in motion the development of Quantum Mechanics through subsequent papers by Born, Jordan, Heisenberg and also Dirac (from 1925 to 1927) is reexamined through a modern lens. In this paper, we shall discuss some new perspectives on (i) what could be {\it the} guiding intuitions for his discoveries and (ii) the origin of the Born-Jordan-Heisenberg canonical quantisation rule. From this vantage point we may get an insight into Einstein's Quantum Riddle \cite{Lande1974,Sommerfeld1918,Born1926} and a possible glimpse of what might come next after the last 100 years of Heisenberg's quantum mechanics.

(This is the first draft of a paper dedicated to the celebration of 100 years of quantum mechanics, on the anniversary of Heisenberg's founding paper on the subject in July 1925, to be published in a celebratory volume in July 2025 by World Scientific Publications, Singapore).

\end{abstract}

\maketitle

\section{Introduction}
\label{Introduction}
Unlike earlier discoveries in theoretical physics prior to 1925, Quantum Mechanics was the result of the efforts of more than one man, namely: Heisenberg, Born, Jordan, Dirac and Schr{\"o}dinger.  It was indeed Heisenberg who made the first and major step in the discovery of Quantum Mechanics, his priority and legacy is therefore unquestionable \cite{Bernstein2005}. However, had he not made that step, it is clear from the historical developments that it would have been Schr{\"o}dinger who would claim priority, though his perspective using the de Broglie matter wave mechanical approach, though ubiquitous, had turned out to be incorrect eventually.  Some physicists famously Weinberg \cite{Weinberg1992} and also Fermi \cite{Fermiref2013}, considered Heisenberg's paper of July 1925 incomprehensible because it was clouded in mystery as to why he made the assumptions that he did and what motivated the steps in his thinking. In 2004  Aitchison et al \cite{Aitchison2004} made an attempt to provide an ``understanding" of Heisenberg's July 1925 paper by filling in the apparent gaps in the calculational details (some omitted by Heisenberg in his 1925 paper) thereby making a reconstruction of this landmark paper. While their effort is instructive and may be useful for the teaching of advanced courses in quantum mechanics, and may have relieved Weinberg's and Fermi's ``incomprehensibility" objection, it has done little in my opinion to elucidate the ``magical" part of Weinberg's observations. Although one concur with Aitchison et al \cite{Aitchison2004} that it may not in fact be possible to ``render completely comprehensible the mysterious processes" whereby physicists ``gain new insights about nature" through a phenomenal breakthrough, nevertheless there are some useful clues that are worth exploring.  This is so, especially as it is now a century after the development of the subject, many of whose stones had been laid as early as the period from 1925 to 1927 and are in need of scrutiny under a modern lens. By doing so we will not only make the process of discovery less ``magical" and plausible, but in this case we may also gain some new insight, one missed during the last century, that might lead  us closer to a solution for Albert Einstein's famous Quantum Riddle, by which he meant: what are ``the principal reasons behind the quanta?" \cite{Lande1974,Pais1982-1991}.
The words Quantum Riddle were coined by Einstein himself ({\it not} Lande \cite{Lande1974}) as early as 1923\cite{Paty1993} but the spirit of the riddle has been in his writings and correspondence even earlier since 1916\cite{Lande1974} until his death in 1954 with numerous interpretations as to its meaning \cite{Paty1993}. For this paper we shall stick closely to Lande's \cite{Lande1974} version, in fact explicitly defining here the quantum riddle as the origin for the Born-Jordan-Heisenberg quantization rule eqn(\ref{BJ Rule}) from which everything else quantum follows \cite{Choy1}. This is the aim in this paper. Even with this limited aim, this paper has found a number of fascinating conclusions and speculations  that will hopefully motivate further research for the next 100 years.

\section{Heisenberg's discoveries - a summary}
\label{HeisenbergSummary}
 The central result of Heisenberg's landmark paper of July 1925 \cite{Heisenberg1925} as later clarified by Max Born and Pascual Jordan \cite{BornJordan1925}, was the quantum rule:
 \begin{equation}
\ [{\hat q},{\hat p}\ ]= i\hbar {\hat 1} \ ,
\label{BJ Rule}
\end{equation}
which Born considered a postulate of the new quantum theory and even today is still accepted by many as such or as an empirical rule beyond that of the old Bohr-Sommerfeld quantization rule. In modern language the left hand side is the famous commutator product of two non-commuting Hermitian operators (aka matrices in the early days) while the right hand side is the imaginary constant $i\hbar$ multiplied by the unit operator.  As we shall see later, one can call the LHS a closure operator.  Heisenberg in his July 1925 paper \cite{Heisenberg1925} was able to obtain only the diagonal part of this expression (not knowing anything about matrices at that time, which was all the more remarkable). Within a matter of days of receiving Heisenberg's paper, Born immediately realised that it represented a matrix product, conjectured that the off-diagonal elements are zero and recruited the assistance of Pascual Jordan to settle the issue (which they did) and produced their more comprehensive paper which formed the foundation of quantum mechanics as we know it today \cite{BornJordan1925}.   Born considered his discovery to be so important and fundamental that he had it engraved on his tomb stone \cite{Bernstein2005,Schoenhammer2025}. It is perhaps interesting to speculate as to the reason why he did this, apart from the fact that Born was obviously very proud of his discovery (which he had said so himself in a commentary of his July 1925 letter to Einstein \cite{Born-Einstein Letters}). Bertrand Russell who contributed a forward to the book \cite{Born-Einstein Letters} had mentioned that both men were ``brilliant and humble", so pride or ego could not have been the answer or the only answer.  A little history may throw some light on this. Archimedes was considered perhaps the greatest scientist and mathematician in antiquity.  He had the result of his favourite theorem that the volume {\it or} surface area of a sphere circumscribed in a cylinder is 2/3 that of the volume {\it or} surface area of the cylinder (first published in 225 BC) made into a sculpture with inscriptions in his tomb stone which was later discovered by the Roman statesman and philosopher Marcus Tullius Cicero (106-43 BC).  Archimedes basically issued a challenge to future generations: find another way of proving this that is better than mine \cite{Archimedes}. As is well known a formal proof was not available until the advent of integral calculus in the 17th century. Max Born in one's opinion issued a similar challenge, for though critical of Einstein's objections to the Copenhagen interpretation in his later years, he was well aware of Einstein's Quantum Riddle and had to remind future generations that it should not be forgotten. Born apparently continued to toy with the fundamental significance of his commutation rule until the war in collaboration with Klaus Fuchs according to Bernstein \cite{Bernstein2005a}.

We are now jumping ahead of the story so let us back track a little. In July 1925 Heisenberg had no idea about matrices, no concept of what we now know as the basic formalism of quantum mechanics.  His ``magical" discovery  essentially came from an assignment set to him by Born, to find this new mechanics which Born already knew must exist as early as 1924, to which he first gave the name Quantum Mechanics \cite{Born1924}. Born also suggested in regular discussions with his two assistants that they focus on the transition amplitudes and look for ``some kind of symbolic multiplication" of these quantities that would form the new mechanics \cite{VdW21} . It is worth reading this paper \cite{Born1924} again to find out why he thought so. Heisenberg too must have studied this paper before he took off to Helgoland.  In addition he had his own paper with Kramers written in January 1925 \cite{HeisenbergJan1925} and his earlier discussions with Bohr on the same subject of dispersions and transition amplitudes \cite{HeisenbergBohr} in the summer of 1922. So by the time he left for Helgoland he had this knowledge with him and his strategy was clear too. Rather than trying out his new ideas on the atomic orbits of the Bohr hydrogen atom model which had caused him difficulties, he chose a simpler model: the one dimensional anharmonic oscillator.  He also knew from his discussions with Bohr that the harmonic oscillator model might be too simple as transitions there are all evenly spaced harmonics just as in the classical case, an issue he had mentioned to Bohr in 1922 in relation to the quadratic Stark effect, which caused Bohr some discomfort \cite{HeisenbergBohr}. On this he was perhaps overly cautious or super prescient (see section \ref{BeyondHeisenberg} later) and he made a fundamental error to postulate a theory based entirely on {\it observable} quantities among which he considered the particle co-ordinates not to be. As noted by van der Warden, this was an error but  a `fruitful error' \cite{VdW33} as it caused him to focus on developing methods for calculating only those quantities which were directly accessible at that time. Heisenberg's theory was developed on the basis of a number of assumptions that we now summarise with some condensation of the ideas and updated notes \cite{VdW28}.

(1) In the atomic range, classical mechanics is no longer valid.

(2) Any new theory must satisfy Bohr's Correspondence principle i.e. in the limit of large quantum numbers, it must agree with classical mechanics. An important device used to satisfy this was the trick of replacing differential quotients in classical formulae by difference quotients.  This trick originates from Born \cite{Born1924} and has been used by Kramers and Heisenberg \cite{Heisenberg1925}.

(3) Heisenberg's hypothesis.  Heisenberg felt that the difficulties encountered at that time were due to the failure of the kinematics underlying the mechanics but {\it not} its laws which remain the same as in the classical theory. Hence the equations of motion such as Newton's law: $\ddot x=f(x)$ should remain the same, except that the kinematic quantity $x$ must be re-interpreted. This was perhaps the single most powerful insight of his paper, it was bold and it was super intuitive and could be considered ``magical".  Later we shall have more to say as to how he could have been so convinced. It remains today as his most fundamental quantization axiom; however a more thorough scrutiny is long overdue.  In one's opinion, too much focus has been given to the Uncertainty Principle and Entanglement in the last 100 years without a thorough examination of this hypothesis which underlies the foundation of quantum mechanics.

(4) Knowing that the classical Fourier series: for a periodic quantity\cite{note1} $x(n,t)=\sum_\alpha a_\alpha(n) e^{i \alpha \omega_n t}$ is only likely to be valid for large quantum numbers he set about to make the transition to quantum quantities by replacing the classical amplitudes $a_\alpha(n)$ by $a_\alpha(n,n-\alpha)$ and the classical resonance frequencies $\omega_n$ by quantum transitional frequencies $\omega(n,n-\alpha)$. This was a major breakthrough on his part, though it had been earlier suggested by Born \cite{Born1924}.

(5) From the assumption (4), Heisenberg then deduced that the quantities concerned have certain `multiplication features'. He showed how this is done for $x^2(t)$ and also $x(t)y(t)$ which is later recognised by Born as matrix multiplication.  To do so, he was guided by the Ritz spectral combination rule: $ \omega(n,n-\alpha)+ \omega(n-\alpha,n-\beta)=\omega(n,n-\beta)$ but had to invoke the Sommerfeld-Wilson quantum rule.  Unfortunately this makes his theory and the later papers by Born, Jordan and himself somewhat ad-hoc and semi-empirical \cite{BornJordan1925} as they were derived essentially from the old quantum theory, in particular Planck's radiation formula. We will return to this point later.

(6) To make further progress he must now obtain a formula to generalise the old quantum rules of Bohr and Sommerfeld. This uses the trick given by (2) above as well as the matrix multiplication rule.  He effectively derived the diagonal elements of the commutator eqn(\ref{BJ Rule}) and,  using a correspondence formula obtained by Born \cite{Born1924}, showed that it is equivalent to the Thomas-Kuhn sum rule of dispersion \cite{Heisenberg1925}.  Heisenberg seemed to intuitively realise that his ``algebra" is incomplete or unspecified without this rule eqn(\ref{BJ Rule}). One can call this his closure hypothesis.

As the details of Heisenberg's paper and other commentaries now exist, we need not repeat them here \cite{Aitchison2004,MacKinnon1977}.  Instead we shall focus on items (3), (5) and (6). However in the seminal paper of 1925 \cite{BornJordan1925}, the Born-Jordan rule eqn(\ref{BJ Rule}) was only discovered as an extension of Heisenberg's earlier work \cite {Heisenberg1925} from the old quantum theory. This was somewhat ad-hoc as mentioned above and was therefore not rooted in very strong theoretical foundations which were, indeed not possible at that time.

Let us first consider item (3) above, as to why Heisenberg felt that it was the kinematics which was at fault in trying to develop a new form of mechanics. First he had by then spent a great deal of time with Bohr and almost everyone at that time was familiar with the success of the Bohr's model for the H atom.  Bohr did not use any fanciful ideas, no new supersymmetric fields or gauge boson field models etc to quote a common present day parlance. Bohr used the Rutherford model and plain old Newtonian mechanics albeit with two new ingredients: the Einstein-Planck energy frequency relation and a postulate for the notorious quantum jumps between stationary states, a concept retained till this day. Next Heisenberg had spent some time talking with Einstein, and he must have convinced himself that something like special relativity was at work here. Einstein told him that it is theory that decides what can be observed, so he decided to pursue a theory based on so called observables.  Now special relativity was a modification of Newtonian theory based on a change of kinematics, not dynamics.  So Heisenberg, then only 23 must also have convinced himself to do the same. So take Newton's law and make some kinematic changes to discover the new mechanics.  The question was what kinematic changes? His knowledge about relativity must be as good as that of his contemporary Wolfgang Pauli, who at the young age of 21 had already written a full monograph on relativity covering both the special and general theories. He might even have tried generalising his observables from vectors to tensors or even dyadics and things like that which he might have found in his textbooks, but failed in the attempt. So he chose to use the Fourier series and knowing the coefficients and Ritz law he came up with non-commuting objects that lead to a spectrum that agreed with experiments, for by that time the vibrational spectrum of atoms and molecules was well known.  In addition he has now a way to calculate transition amplitudes. Could he have done better?  One may think the answer to this question is yes, if he was familiar with the French literature \cite{PrivateComm1}.  The reason is that between the time of Newton and Einstein, a most significant theory of dynamics had been found which has deep implications to the present day.  This was the treatise: Traite de Dynamique \cite{dAlembert1743} published by the French physicist and mathematician Jean-Baptiste le Rond d'Alembert who preceded Hamilton and Lagrange by defining the force of inertia and incorporating it into the principle of virtual work and so extending it from statics to dynamics. Pars \cite{Pars} called d'Alembert's principle  the fundamental law of mechanics, superseding the laws of Newton. The implications of d'Alembert's principle for mechanics and especially quantum mechanics are deeply profound.  Unfortunately for the last 100 years this connection has rarely been made or fully explored.  For d'Alembert, dynamical motion is an equilibrium condition once we incorporate the force of inertia.  Now Heisenberg could have better understood why he must introduce new kinematic objects in quantum mechanics, if he had realised that it is required by d'Alembert's principle. In classical mechanics, d'Alembert's principle can be stated in terms of the applied force and the inertial force in an equilibrium condition using the principle of virtual work, {\it subject to harmony with given kinematic constraints} \cite{Lanczos1986}. For N bodies this is usually written in the form:
\begin{equation}
\sum_{n=1}^{N}({\bf F}_n-m_n {\bf A}_n)\cdot \delta{\bf R}_n=0  \ ,
\label{dAlembert1}
\end{equation}
where ${\bf F}_n$ is the external applied force, ${\bf A}_n = {\bf \ddot R}_n$ is the acceleration and $\delta{\bf R}_n$ is the virtual displacement of each particle $n$.
So what Heisenberg is intuitively saying is that d'Alembert's principle in the form eqn(\ref{dAlembert1}) is untenable.  He could and may have tried other extensions of the equation to tensors, dyadic, quaternion and other objects.  In the modern context most of these attempts would have been fruitless if he had indeed made such attempts.  This is because they are inadvertently local hidden variable theories that we now know are doomed to failure.  Instead Heisenberg opted to choose his quantisation procedure, in the non-relativistic case, by an extension to objects which he later found out through Born are some kind of matrices, nowadays known to be in fact complex Hermitian operators.  So we put hats on all the old classical quantities to indicate this:
\begin{equation}
\sum_{n=1}^{N}({\bf \hat F}_n-m_n {\bf \hat A}_n)\cdot \delta{\bf \hat R}_n=0  \ .
\label{dAlembert2}
\end{equation}
This is a deep and powerful statement, it says that with these objects, d'Alembert's equilibrium principle is once again restored.  That, one must think is {\it the fundamental postulate} of quantum mechanics. While we could find no direct evidence that Heisenberg's intuition was guided by d'Alembert's principle, it is not inconceivable that he had been influenced by Sommerfeld's lectures and therefore his ``magical" hypothesis item (3) was something quite natural for him.  Obviously with such a view the measurement postulates that Dirac later had to propose and von Neumann adopted, including the difficulties with concepts of quantum reality of EPR, wavefunction collapse and so on could have been quite different, if not avoidable. Naturally one can pursue this quantum d'Alembert's equilibrium principle even more deeply. For sure one should not think it will lead us back to the classical equilibrium conditions d'Alembert envisaged at his time and in the words of Einstein `` the Lord is subtle", but with equilibrium one will immediately see fluctuations, see later.   To pursue this Class I(a) problem \cite{Choy1} would take us too far afield so for now let us return to points (5) and (6).  Having decided that he must create new quantum objects which he had shown to be non-Commutative, which greatly disturbed him, he must have realised that the algebra is incomplete and worse even inconsistent without setting up some rules. Consider a simple example: let C=AB and D=BA. Then Y=ABA can be written as Y=CA or Y=AD.  To be consistent clearly this requires CA=AD and similarly  DB=BC. As a matter of fact, in 1928 Hermann Weyl caught on to this and to avoid other issues with the Born-Jordan-Heisenberg quantisation rule eqn(\ref{BJ Rule}), such as the trace anomaly \cite{Choy2020a}, he solved this consistency issue by proposing the quantisation rule: $AB=e^{i\alpha}BA$, since named after him, with A and B being unitary operators and $\alpha$ a real number phase factor \cite{Weyl1931}.  Heisenberg's concerns must have disturbed him so much that he went about finding out how to do the algebra using the technique as mentioned earlier of Born's in item(4) above, finishing with the diagonal part of the quantum rule eqn(\ref{BJ Rule}) via items (5) and (6). As noted earlier this approach is ad-hoc but his procedure worked and the results are in agreement with experiment though it may appear to be magical or even mysterious.   Einstein for one complained that `` the theory says a lot but does not really bring us any closer to the secret of the `old one' " \cite{Born-Einstein Letters2}.  For the rest of this paper, we shall show how Heisenberg's motivations can be put on a more fundamental basis, that may perhaps have been more palatable to Einstein.

\section{A Re-Foundation of Quantum Mechanics - Dirac Quantization}
\label{Refoundation}
As d'Alembert's principle involves an inertial force that is of a polygenic character, the usual practice is to perform an integration by parts \cite{Lanczos1986-2} and convert d'Alembert's variational principle into Hamilton's least action principle involving only scalar functions. Note that this process is only valid classically for monogenic applied forces and holonomic constraints \cite{Lanczos1986}.  We shall assume the same after quantisation. Note also it is only after this process is carried out in classical mechanics that the concepts of virtual and actual paths for the particles become possible.  In configuration space the latter can be shown to be determined by Jacobi's principle \cite{Lanczos1986}, of great importance in Schr\"odinger's Wave Mechanics.  Clearly in the quantisation of the d'Alembert's principle as presented here, these concepts now no longer have the same meaning. Nevertheless when Born and Jordan \cite{BornJordan1925} realised that the objects Heisenberg had introduced were matrices they did try to propose a variational principle to justify the quantum Hamilton equations in their analysis, which was unfortunately severely inadequate and subsequently ignored in their three men paper \cite{BJH1926}.  Therefore eqn(\ref{BJ Rule}) remains an ad-hoc rule which Born had always considered a postulate \cite{Bernstein2005,MaxBornNobelLectures,Schoenhammer2025}, although not explicitly considered as an axiom in von-Neumann's axiomatic foundations.  It was left to Dirac \cite{Gottfried2011,Dirac1926} who in 1926 tried to lay a more fundamental theoretical foundation by proposing his more general quantum Poisson bracket quantisation condition from first principles. However Dirac, in spite of being a native French speaker, was unlikely to have been influenced by d'Alembert's principle; he never cited d'Alembert's principle in any of his works.  He was a Hamiltonian man, being influenced in his own words, by Whittaker's ``Analytical Dynamics" which contained no mention of d'Alembert's principle\cite{Whittaker1917, Dirac1961}.  Dirac's rule was:
\begin{equation}
 \{{\hat u},{\hat v}\}=-\frac{i}{\hbar}[{\hat u},{\hat v}] \\  \ ,
\label{Dirac Rule}
\end{equation}
where the operator Poisson bracket $\{{\hat u},{\hat v} \}$ on the LHS must be defined as:

\begin{equation}
\{{\hat u },{\hat v}\}=\bigl (\frac{\partial \hat u}{ \partial \hat q} \circ  \frac{\partial \hat v}{ \partial \hat p}  - \frac{\partial \hat u}{ \partial \hat p} \circ \frac{\partial \hat v }{\partial \hat q}\bigr )\ , \\
\label{P1definition}
\end{equation}
by which the Born-Jordan quantisation rule eqn(\ref{BJ Rule}) is a corollary, (taking ${\hat u }={\hat q }$ and ${\hat v }={\hat p }$) .  Here we retain the $\circ$ to denote a symmetrised product, (ignored by Dirac),  whose details we will not go into here \cite{Zoran2003}, except that it is symmetric in the exchange of the multiplicands. The Dirac rule eqn(\ref{Dirac Rule}) was initially obtained through correspondence arguments in the case of large quantum numbers \cite{Gottfried2011,Dirac1926}. He was obviously dissatisfied with this approach that he used to propose his quantum law, since in his seminal book of 1931 he proposed a separate theoretical argument, that detached the correspondence principle. Moreover he was unclear what eqn(\ref{Dirac Rule}) actually meant.  Rather than a mathematical equality he seemed to later settle on a corresponding identity or a replacement rule or algorithm for quantisation. Nevertheless this was a fundamentally important advance and his argument was ingenious.  Any student of his classic book\cite{Dirac1931} should be familiar with it and it was based on algebraic manipulations of a four operator Poisson bracket using the Leibniz rule, nowadays associated with dynamical Lie algebraic group properties \cite{Dirac1931}. Appendix 1 also gives an alternative proof of this important rule.  With this ``derivation", Dirac canonised his quantum condition eqn(\ref{Dirac Rule})  as the foundation of quantum mechanics and then went on to show that the Born-Jordan rule eqn(\ref{BJ Rule}) can be obtained from it. In 1961, in a letter to van der Warden, Dirac confessed that he had obtained his condition in 1926 first because he had expected ``some kind of connection between the new mechanics and Hamilton mechanics".  It is unclear however why Dirac was satisfied with his 1931 derivation of his rule \cite{Dirac1931} because its connection between quantum mechanics and Hamiltonian mechanics contains some logical gaps, which were never really clarified in 1931 nor in fact up to the present day. Perhaps he had searched all his life and failed, or perhaps he felt the problem is sufficiently solved so that he felt no longer bothered by it. In particular, starting from Poisson brackets, the Leibniz rule provides logically only a one way proof of the relation: $ \{{\hat u},{\hat v} \} \Rightarrow = \frac {1}{i\hbar} [{\hat u},{\hat v}]$ . Owing to this he continued to discuss this rule as a correspondence rule or quantisation rule but not as an equality, let alone an identity.  Closer examination shows that his ``derivation"  can also be viewed as a consistency condition between multi operator quantum Poisson bracket algebra with the Lie algebra of quantum commutators and thus cannot on its own constitute a fundamentally complete derivation of canonical quantisation \cite{Dirac1931}, see Appendix 1. We shall later also discuss other consistency conditions by which we can extend our arguments to quantum Lagrange brackets, a subject that appeared to have been missed in the quantum literature for nearly a hundred years. Later authors such as Groenewold \cite{Groenewold1946} also discovered inconsistencies in Dirac's scheme which nowadays go by the name of ``quantization obstructions", technical difficulties we shall not go into in this work.

In this section we shall first fill the logical gaps in Dirac's canonical quantisation scheme and show that it can actually be derived in the reversed way, i.e. $ [{\hat u},{\hat v}] \Rightarrow i\hbar \{{\hat u},{\hat v} \}$. In his 1963 Scientific American article, Paul Dirac \cite{Dirac1963} proposed as a major class 1 problem the question: `` How can one form a consistent picture behind the rules for the present quantum theory?", perhaps his own way of framing Einstein's Quantum Riddle.  Understanding the origin of these rules may help put us closer to the ``consistent picture" he so desired. Recall that in classical mechanics, Hamilton's variational principle requires that {\it both} p and q can and must be varied independently to obtain the Hamilton equations of motion,(see for example Lanczos \cite{Lanczos1986-3} ). This procedure is a unique consequence of the Legendre transformation that takes one from the Lagrangian $L$ to the Hamiltonian function $H=p{\dot q}-L$. In quantum mechanics this independence of variation is no longer  possible due to non-commutativity and the requirement of rules to define the algebra.   To restore this independence, we shall now write (restricting to one dimension without loss of generality) the quantum action as:
\begin{equation}
{\hat S}= \int \Bigl({\hat p} d {\hat q} - {\hat H}({\hat p},{\hat q}) dt - \lambda {\hat F}({\hat u},{\hat v}) dt \Bigr ). \\
\label{QMAction1}
\end{equation}
Strictly speaking a term like ${\hat p} d {\hat q}$ should be replaced by its symmetrised form $\frac{1}{2}({\hat p}\ d {\hat q}+ d {\hat q}\ {\hat p})$ and there will be ordering issues in defining the Hamiltonian ${\hat H}({\hat p},{\hat q})$ which however will not affect our arguments here \cite{Zoran2003}. The kinematic constraint function ${\hat F}({\hat u},{\hat v})$ will be further discussed below, where $\hat u(\hat p,\hat q)$ and $\hat v(\hat p,\hat q)$ are arbitrary functions at this stage. We must emphasize here that as soon as the classical Hamiltonian action is replaced by the quantum one with operators then due to non-commutativity a relationship must exist between the operators as mentioned earlier to ensure consistency and algebraic closure \cite{Dirac1931a}.  The variation is no longer constraint free, even for a non-interacting particle as in the classical case.  Here $\lambda ({\hat u},{\hat v}) $ is the Lagrange multiplier and we have assumed it as a function of ${\hat u}$ and ${\hat v}$ for simplicity,  but in fact it need not be \cite{DiracLectures}.  The reader familiar with the theory of Lagrange multipliers will recall that in general $\lambda$ need not be determined, hence the name undetermined multiplier and in the case of a vanishing derivative of the constraint (see below), it is in fact undeterminable.     Now there are in fact many ways to choose the function ${\hat F}({\hat u},{\hat v})$ that will satisfy the Heisenberg hypothesis item (3) as we shall see.  One way as alluded to by Dirac \cite{Dirac1931a}, is that quantisation can be specified either by a rule using either the commutator or the anti-commutator. Another is the Weyl rule as discussed above \cite{Weyl1931}. Although ${\hat u}({\hat p},{\hat q})$ and ${\hat v}({\hat p},{\hat q})$ are arbitrary functions at the moment, later their specific choice will determine the quantisation rule.  The inclusion of a constraint function ${\hat F}({\hat u},{\hat v})$ in the action is what's missing in the original variational principle of Born and Jordan's paper in 1925 \cite{BornJordan1925}. Also why was the specific choice  ${\hat u}={\hat q}, {\hat v}={\hat p}$ necessary as a quantization rule in eqn(\ref{Dirac Rule}) was never given by Dirac \cite{Dirac1931} (1931), other than that it yields the corollary Born-Jordan rule eqn(\ref{BJ Rule}). Once again, note that the algebra cannot be closed or consistent, without specifying the constraint and any application of the variational principle must respect that,; see statement in italics before eqn(\ref{dAlembert1}).  So what can Heisenberg's hypothesis (see item (3) above) that Newton's law will continue to hold, tell us about the unknown constraint function ${\hat F}({\hat u},{\hat v})$. Is it always true, or indeed was Heisenberg just a lucky man?  As can be easily seen, a straightforward application of the variational calculus will now provide us with the equations of motion (see for example Landau and Lifshitz \cite{Landau1} ):
\begin{equation}
{\dot{\hat q}}= \frac{{\partial \hat H}}{\partial \hat p} + \lambda \frac{{\partial \hat F}}{\partial \hat p}\ {\rm and}\ {\dot{\hat p}}= -\Bigl ( \frac{{\partial \hat H}}{\partial \hat q} + \lambda \frac{{\partial \hat F}}{\partial \hat q}\ \Bigr ).  \\
\label{HamiltonEqns1}
\end{equation}
For Heisenberg's hypothesis to be valid, the canonical equations of motion must remain unchanged.  Therefore we require:
\begin{equation}
 \lambda \frac{{\partial \hat F}}{\partial \hat p}\ =0\  and \  \lambda \frac{{\partial \hat F}}{\partial \hat q}\ = 0.  \\
\label{HamiltonEqns2}
\end{equation}
However there are many ways in which this can be satisfied. As an example consider the Weyl quantisation rule \cite{Weyl1931} in which the constraint function ${\hat F}({\hat u},{\hat v})$ now takes the form:
\begin{equation}
{\hat F}({\hat q},{\hat p})=\ {e^{is\hat q}} {e^{it\hat p}}\ - e^{-i\hbar s t} {e^{it\hat p}} {e^{is\hat q}}=0,
\label{Weyl Form}
\end{equation}
in which eqn(\ref{BJ Rule}) must still hold implicitly for it to be valid.  Weyl's rule is nowadays better known as a special case of the Baker–Campbell–Hausdorff formula, which has a long history \cite{Baker-Campbell2012}.  Here it is straightforward using Leibniz's rule to show that eqn(\ref{HamiltonEqns2}) is now satisfied by way of the partial derivatives vanishing. However we must be clear that this is not the object of our exercise, that is to insert the known quantum rules as the constraint function and derive quantum mechanics. This would at best only give us a bootstrap theory. Instead we want to exploit the constraint function as a means to find out more about Albert Einstein's Quantum Riddle \cite{Lande1974} or in his own words: ``If only I knew which little screw the Lord applies here" \cite{Sommerfeld1918}.
An alternative could also be that $\lambda$ is in fact zero in which case the derivatives of $F$ do not matter. We shall look further into all these factors later; for now we shall follow Dirac's \cite{Dirac1931a} specification and assume that the constraint can be given as some unknown function of the commutator alone (plus perturbations which we shall discuss later, and in view of Heisenberg's hypothesis {\it must} be {\it very small} ) i.e. ${\hat F}([{\hat u},{\hat v}])$, see later.   This in turn requires:
\begin{equation}
 \lambda \hat F^\prime  \frac{{\partial}}{\partial \hat p} [{\hat u},{\hat v}] \ =0,  \\
\label{HamiltonEqns3}
\end{equation}
and:
\begin{equation}
 \lambda \hat F^\prime  \frac{{\partial}}{\partial \hat q} [{\hat u},{\hat v}] \ =0.  \\
\label{HamiltonEqns4}
\end{equation}
For $\lambda \ne 0$ (undetermined multiplier) and $\hat F^\prime \ne 0$ i.e. differentiability, we now have:
\begin{equation}
\frac{{\partial}}{\partial \hat p} [{\hat u},{\hat v}] \ =0,  \\
\label{HamiltonEqns5}
\end{equation}
and:
\begin{equation}
\frac{{\partial}}{\partial \hat q} [{\hat u},{\hat v}] \ =0.  \\
\label{HamiltonEqns6}
\end{equation}
Now eqn(\ref{HamiltonEqns5}) and eqn(\ref{HamiltonEqns6}) imply that for the specified choice of ${\hat u}$ and ${\hat v}$ that we must take to define the quantisation rule, the commutator $[{\hat u},{\hat v}]$   cannot be a function of ${\hat p}$ nor of  ${\hat q}$.
It must be a constant matrix operator, but we have not yet proven that it must be diagonal, nor what specified choice to make for ${\hat u}$ and ${\hat v}$ .  To proceed further we need to invoke the Poisson algebra eqn(\ref{P1definition}). Then we must have for any arbitrary ${\hat u}$ and ${\hat v}$\ :
\begin{equation}
\frac{{\partial}}{\partial \hat p} [{\hat u},{\hat v}] \ = \{{\hat q},[{\hat u},{\hat v}]\} \ =0,  \\
\label{HamiltonEqns7}
\end{equation}
and:
\begin{equation}
\frac{{\partial}}{\partial \hat q} [{\hat u},{\hat v}]\ = -\{{\hat p},[{\hat u},{\hat v}]\} \ =0.  \\
\label{HamiltonEqns8}
\end{equation}
Now we are nearly there, for under any canonical transformation that takes ${\hat p} \rightarrow {\hat P}$ and ${\hat q} \rightarrow {\hat Q}$ {\it and} must leave the canonical equations of motion eqn(\ref{HamiltonEqns1}) intact, {\it and} indeed also eqn(\ref{HamiltonEqns7}) and eqn(\ref{HamiltonEqns8}), we require that the brackets:
\begin{equation}
\{{\hat q},[{\hat u},{\hat v}]\}_{\hat p,\hat q}= \{{\hat Q},[{\hat u},{\hat v}]\}_{\hat P,\hat Q}  \\
\label{QuantumBrakets1}
\end{equation}
and
\begin{equation}
\{{\hat p},[{\hat u},{\hat v}]\}_{\hat p,\hat q}= \{{\hat P},[{\hat u},{\hat v}]\}_{\hat P,\hat Q} . \\
\label{QuantumBrakets2}
\end{equation}
Here I use the notation of Goldstein \cite{Goldstein1980} where the subscripts indicate the canonical variables associated with the bracket.
Eqn(\ref{QuantumBrakets2}) implies that we must have the necessary condition (Lemma 1):
\begin{equation}
[{\hat u},{\hat v}]_{\hat p,\hat q}= [{\hat u},{\hat v}]_{\hat P,\hat Q}.  \\
\label{QuantumBrakets3}
\end{equation}
There are only two brackets with this property namely Poisson $\{u,v\}$ or Lagrange $(u,v)$ \cite{Goldstein1980}. However because the commutators satisfy the Jacobi identity, as do the Poisson brackets (see Appendix 1) while the Lagrange brackets do not, we finally arrive at Dirac's Quantum Condition see eqn(\ref{QuantumBrakets4}) below. Note that this is not yet a quantisation rule which is why we must be careful to distinguish between what is a quantum condition and what is a quantum rule.  The equation:
\begin{equation}
[{\hat u},{\hat v}]= \kappa \{{\hat u},{\hat v}\},  \\
\label{QuantumBrakets4}
\end{equation}
where here $\kappa= i\hbar$ by dimensional analysis for Hermitian operators, {\it is} a quantum condition. Now since this proof of equality goes in the direction: $[{\hat u},{\hat v}]\rightarrow \kappa \{{\hat u},{\hat v}\}$ i.e. it is a proof of necessity. We also need Dirac's \cite{Dirac1931} proof (see also Appendix 1) which goes in the opposite direction $[{\hat u},{\hat v}]\leftarrow \kappa \{{\hat u},{\hat v}\}$, as a proof of sufficiency.   We can now conclude that Dirac's famous Quantum Condition \cite{Dirac1931} is in fact an identity or equivalence:
\begin{equation}
[{\hat u},{\hat v}] \equiv \kappa \{{\hat u},{\hat v}\}.  \\
\label{QuantumBrakets5}
\end{equation}
Mathematically what we have proven here is that the commutator bracket Lie algebra and the Dirac-Poisson bracket algebra (subject to operator ordering requirements which we cannot go into here) are isomorphisms. That is not all.  Further to eqn(\ref{HamiltonEqns7}) and eqn(\ref{HamiltonEqns8}) and comments following them, there can be only {\it one} choice for the unique quantisation rule, namely: ${\hat u}={\hat q}$ and ${\hat v}={\hat p}$, or vice versa, that satisfies Heisenberg's hypothesis (item 3).  Thus we arrive at the non-commuting canonical Born-Jordan-Heisenberg quantisation rule eqn(\ref{BJ Rule}), as the fundamental rule of canonical quantisation. Note that the commutating relations $[{\hat q},{\hat q}\ ]=[{\hat p},{\hat p}\ ]=0$ corresponding to the choices ${\hat u}={\hat q}$ , ${\hat v}={\hat q}$ and ${\hat u}={\hat p}$ , ${\hat v}={\hat p}$ respectively also follow from our arguments and are trivial here. Note also that in evaluating the Poisson bracket to obtain the quantum rule eqn(\ref{BJ Rule}) using eqn(\ref{QuantumBrakets5}), the latter now collapses into an equality, in fact a weak equality $\approx$ and becomes a secondary constraint, in the terminology of Dirac \cite{DiracLectures}.  It is impossible to prove eqn(\ref{BJ Rule}) from right to left; physically such a general statement would also be absurd. However this does not diminish its status as a fundamental physical law, with tribute to Max Born who recognised it straightway in 1925.

Before moving on, we shall merely state that it is straightforward to generalise the above results to the case of boson field theory.  The standard approach well known since 1929, see for example Heisenberg \cite{Heisenberg1929} and later Schiff \cite{Schiff1949}, Dirac \cite{DiracLectures}, is to generalise all derivatives to functional derivatives, and the quantisation rule for fields becomes:
\begin{equation}
[{\hat \psi}(x),{\hat \pi}(x')] = \kappa \ \delta(x-x'),  \\
\label{FieldQMrule}
\end{equation}
where $\kappa$ is in general a complex dimensional constant (= $i\hbar$ for bosons) and
 ${\hat \pi}(x)$ is the canonical field momentum operator, all other commutators between canonical field variables being zero.  This is usually written in the standard form:
\begin{equation}
[{\hat \psi}(x),{\hat \psi^\dag}(x')] =  \ \delta(x-x'),  \\
\label{FieldQMrule1}
\end{equation}
using the following result: ${\hat \pi}(x)=i\hbar {\hat \psi^\dag}(x)$  from the Lagrangian field density \cite{Heisenberg1929,Schiff1949}, see also eqn(\ref{Appendix3-4}) below.

\section{Lagrange Bracket quantisation}

In this section we shall replace the constraint function ${\hat F }$ in eqn(\ref{QMAction1}) by a function ${\hat G}$ which is explicitly written as a function of an anti-commutator.
\begin{equation}
{\hat S}= \int \Bigl({\hat p} d {\hat q} - {\hat H}({\hat p},{\hat q}) dt - \lambda {\hat G}([{\hat u},{\hat v}]_+) dt \Bigr ), \\
\label{QMAction2}
\end{equation}
where $[{\hat u},{\hat v}]_+={\hat u}{\hat v}+{\hat v}{\hat u}$.  Then all our previous arguments follow up to eqn(\ref{QuantumBrakets3}), by replacing commutators with anti-commutators.  Since anti-commutators and also the Lagrange brackets do not satisfy the Jacobi identity, we must now modify the Dirac Quantum Condition as:
\begin{equation}
[{\hat u},{\hat v}]_+ = {\bar \kappa} P_{{\hat u}{\hat v}} ({\hat u},{\hat v}),  \\
\label{QuantumBrakets6}
\end{equation}
where ${\bar \kappa}$ is a different complex dimensional constant that now depends on the choice for ${\hat u}$ and ${\hat v}$ and $P_{ab}=1$ for the identity permutation of a,b and -1 for odd ones and $({\hat u},{\hat v})$ is the Lagrange bracket:
\begin{equation}
({\hat u },{\hat v})=\bigl (\frac{\partial \hat q}{ \partial \hat u} \circ \frac{\partial \hat p}{ \partial \hat v}  - \frac{\partial \hat p}{ \partial \hat u} \circ \frac{\partial \hat q }{\partial \hat v}\bigr )\ . \\
\label{Lagrangedefinition}
\end{equation}
The factor $P_{ab}$ is required here as the Lagrange bracket $(u,v)$ is antisymmetric in the exchange of $u\leftrightarrow v$ while the anti-commutator is symmetric. The one way proof $[{\hat u},{\hat v}]_+\rightarrow \kappa P_{{\hat u}{\hat v}} ({\hat u},{\hat v})$ is now obtained in a similar way. However, the reverse proof $[{\hat u},{\hat v}]_+\leftarrow \kappa P_{{\hat u}{\hat v}} ({\hat u},{\hat v})$ is less straightforward, Lagrange brackets do not satisfy Leibniz's rule, and Dirac's arguments (see Appendix 1) do not apply. The proof requires super Jacobi identities and it is provided in the Appendix 2.  Eqn(\ref{QuantumBrakets5}) now becomes a new isomorphism:
\begin{equation}
[{\hat u},{\hat v}]_+ \equiv {\bar \kappa} P_{{\hat u}{\hat v}} ({\hat u},{\hat v}).  \\
\label{QuantumBrakets7}
\end{equation}
Eqn(\ref{QuantumBrakets7}) is a fundamentally new result of this paper; it modifies the Born-Jordan-Heisenberg rule eqn(\ref{BJ Rule}) to:
\begin{equation}
\ [{\hat q},{\hat p}\ ]_+ = \hbar {\hat 1} \ .
\label{Choy Rule}
\end{equation}
There is now no imaginary $i$ in this quantum rule as the anti-commutator of two Hermitian operators is Hermitian . There does not seem to be any application for this rule in regular physical systems of particles as far as I am aware.  However, extending to field theory it can be shown to provide a direct justification for the Jordan and Wigner's 1928 hypothesis \cite{Jordan-Wigner1928} for fermions in quantum field theory. Lagrange bracket quantisation provides a fundamental theoretical basis for the construction of a quantum field theory of fermions, ending a near century old puzzle for fermion fields as to the origin of their anti-commutation relations, see Appendix 3.  We can now further state a theorem regarding the mixing of brackets:\\

\noindent Theorem I:

Just as in classical mechanics, the mixing of Poission and Lagrange brackets in a mathematical expression, which upon quantisation become commutators and anti-commutators is allowed, see Appendix 3.  However mixing of quantization rules is forbidden.\\

\noindent A simple counter example suffices.  Consider again ABC \cite{Dirac1931a} and assuming this is allowed i.e. $ABC=[A,B]C+BCA= i\hbar C+BCA$ using the quantum rule eqn(\ref{BJ Rule}). Then also $ABC=[A,B]_+C-BCA= {\bar \kappa} C-BCA$ by the quantum rule eqn(\ref{Choy Rule}). By simple manipulation we have: $2 BCA=({\bar \kappa}-i\hbar) C$.  Now taking determinants and using $det(ABC) = det A\ det B\ detC$ we now have $2 det A\ det B= ({\bar \kappa}-i\hbar)$.  For Hermitian matrices, the LHS is always real while the RHS is complex, ${\bar \kappa}$ being real. We have a contradiction, hence by {\it reductio ad absurdum} the theorem is proved.

To conclude this section, we shall make some remarks about Lagrange brackets.  Lagrange brackets have been historically displaced by Poisson brackets in quantum mechanics. In celestial mechanics and astronomy, at least in the days before electronic computing, Lagrange brackets are of far more value, especially in perturbational calculations \cite{Smart1953}. A revival of Lagrange brackets in quantum mechanics would be of great importance. Lagrange brackets and therefore anti-commutators do not form a Lie Algebra.

\section{Beyond Heisenberg}
\label{BeyondHeisenberg}
Let us start with a quick review of Heisenberg's quantum mechanics in terms of Poisson operator algebra.  We first write down the total Hamiltonian ${\hat H}_T$ (symbolically) as :
\begin{equation}
{\hat H}_T= {\hat H} + {\hat \lambda}\circ{\hat F}, \
\label{Hamiltonian1}
\end{equation}
where ${\hat F}$ is the mysterious `` Quantum Red October" function \cite{RedOctober} (for want of a better name)  and ${\hat \lambda}$ is the equally illusive (undetermined) Lagrange multiplier operator, which we shall call its ghost. The time evolution of any dynamical operator $\hat G$ is given as usual by:
\begin{equation}
\dot {\hat G }= \{\hat G,{\hat H}_T\} = \{\hat G,{\hat H}\}+\{\hat G,{\hat \lambda}\circ{\hat F}\}= \{\hat G,{\hat H}\}+{\hat \lambda}\circ \{\hat G,{\hat F}\}+ \{\hat G,{\hat \lambda}\}\circ {\hat F} . \
\label{ReadOctober1}
\end{equation}
We leave the  symbol $\circ$ as a multiplier to remind the reader that in general a symmetrization procedure is required in order that Leibniz's rule can be used in the second equation to expand the Poisson brackets above, although we will not consider those exotic Hamiltonian systems where such a procedure is required in this paper \cite{Zoran2003,Choy2020a}. The last term in eqn(\ref{ReadOctober1}) can now be dropped.  Now Heisenberg's hypothesis:
\begin{equation}
\dot {\hat G }= \{\hat G,{\hat H}\}  \
\label{ReadOctober2}
\end{equation}
is equivalent to:
\begin{equation}
{\hat \lambda}= 0 \  {\it or}\ \{\hat G,{\hat F}\}=0 . \
\label{Heisenberg Hypothesis}
\end{equation}
In particular if the constraint is time independent, which implies that the quantum rule such as eqn(\ref{BJ Rule}) is also time-independent, then we must have $\dot {\hat F }=0$ which requires from eqn(\ref{ReadOctober1}) that $\{\hat F,{\hat H}\}=0$ i.e, both $\hat F$ and $\hat H$ must commute. In this case we must also have:
\begin{equation}
\dot {\hat \lambda}= \{\hat \lambda,{\hat H}\}+{\hat \lambda}\circ \{\hat \lambda,{\hat F}\}=0 \Rightarrow \{\hat \lambda,{\hat H}\}=0\ {\it and}\  \{{\hat \lambda},{\hat F}\}=0,  \
\label{RedOctober3}
\end{equation}
by eqn(\ref{ReadOctober1}). The last requirement is redundant if the operators are (i) Hermitian and (ii) non-degenerate, but we cannot assume that, as commutativity is in general non transitive.  ${\hat \lambda}$ and ${\hat F}$  can both be Hermitian i.e. are observables or anti-Hermitian and therefore are not observables, but the Hamiltonian ${\hat H}$ and total Hamiltonian ${\hat H}_T$  must be Hermitian.  Nevertheless, we can now define a class of operators that satisfy Heisenberg's hypothesis, which we shall call the Heisenberg class.  For a given Hamiltonian $\hat H$, the set of mutually commuting operators $\hat H$, $\hat \lambda_j$ and $\hat F_j$, $j=1,2,3...$ will define a (Heisenberg) class of quantum mechanical systems, all having the same equation of motion albeit with apparently different quantum rules such as eqn(\ref{BJ Rule}) and eqn(\ref{Weyl Form}). In short there are many $\hat F_j$ functions and many ghosts. However for the same initial conditions i.e. a prior prepared state, the evolved states are therefore all identical. We shall not examine the details of this algebra further here, which is rather rich.  For example, it is easy to see by successive evolutions using a pair ${\hat \lambda}_i$, ${\hat F}_i$ and another pair ${\hat \lambda}_j$, ${\hat F}_j$ that we have (see Appendix 4):
\begin{equation}
\Delta^2 \hat G=\delta \hat G'- \delta\hat G''=\hat\gamma_i\hat\epsilon_j\{\hat G,\{\hat F_i,\hat F_j\}\}+\{\hat\gamma_j,\hat\epsilon_i\}\{\{\hat G,\hat F_j\},\hat F_i\}.
\label{RedOctober3a}
\end{equation}
Hence unless the $\hat\lambda$ ghost algebra is commutative i.e. $\{\hat\gamma_j,\hat\epsilon_i\}=0$, the $\hat F_j$ do not even form a group i.e. $\{\hat F_i,\hat F_j\}=\hat F_l$. Eqn(\ref{RedOctober3a}) vanishes for the Heisenberg class which forms a commutative group.   As noted earlier this class is commutative, hence its members are first class primary constraints \cite{Dirac1961} and do not lead to changes of state. This appears to be related to Herman Weyl's\cite{Weyl1931} observation way back in 1928 that the quantum rule eqn(\ref{BJ Rule}) forms an Abelian group in {\it ray} space.

For now we shall merely look at the evolution of the quantum bracket ${\hat \chi}=\lbrack\!\lbrack {\hat u},{\hat v}\rbrack\!\rbrack$ which could be the commutator or anti-commutator.  From eqn(\ref{ReadOctober1}) we have:
\begin{equation}
\dot {\hat \chi}= \{\hat \chi,{\hat H}\}+{\hat \lambda}\circ \{{\hat \chi},{\hat F}\}=0,
\label{RedOctober4}
\end{equation}
for time-independent evolution of the bracket. Clearly as discussed earlier in section \ref{Refoundation}, Heisenberg's hypothesis is equivalent to $\{\hat \chi,{\hat H}\}=0$ {\it and} $\{\hat \chi,{\hat F}\}=0$. The latter is true {\it if} $F$ is only a function of $\hat \chi$ as assumed in section \ref{Refoundation}. Failure of eqn(\ref{RedOctober4}) will lead to perturbations and $\hat \chi$ is no longer a constant of motion, with consequences for entanglement and the uncertainty principle.

\subsection{Schwinger's quantum rule}
The following discussions should be read as symbolic only, as there are some complex mathematical issues that we cannot discuss here.  First we shall make a Legendre transformation back to the Lagrangian:
\begin{equation}
{\hat L}= {\hat p}\dot {\hat q}-{\hat H}_T={\hat p}\dot {\hat q}-{\hat H}-{\hat \lambda}\circ{\hat F}={\hat L}_0-{\hat \lambda}\circ{\hat F},
\label{Schwinger1}
\end{equation}
with ${\hat L}_0$ the unconstrained Lagrangian. Now applying the variational principle to the action ${\hat S}=\int {\hat L} dt$ we now have:
\begin{equation}
\delta {\hat S}= 0 \Rightarrow \delta {\hat S}_0= \delta {\hat W},
\label{Schwinger2}
\end{equation}
where ${\hat S}_0=\int {\hat L}_0 dt$ and ${\hat W}=\int {(\hat \lambda}\circ{\hat F}) dt$.  Eqn(\ref{Schwinger2}) is an extension of Schwinger's variation principle \cite{Schwinger2000}. To see this, consider calculating the expectation value of an operator ${\hat \Omega}$ from eqn(\ref{Schwinger2}):
\begin{equation}
<{\hat \Omega}\delta {\hat S}_0>= <{\hat \Omega}\delta {\hat W}>=-<\delta {\hat \Omega}{\hat W}>,
\label{Schwinger3}
\end{equation}
where the latter follows via a (symbolic) integration by parts, or a variational principle: $\delta <{\hat \Omega}{\hat W}>=<{\delta \hat \Omega}{\hat W}>+<{\hat \Omega}\delta {\hat W}>=0$ . Now Schwinger's variational principle can be written as: $<\delta {\hat \Omega}>=-\frac{i}{\hbar}<{\hat \Omega}\delta {\hat S}_0>$, a result that can also be obtained by Feynman's path integrals .  Eqn(\ref{Schwinger3}) identifies ${\hat W}$ as similar to a Quantum Anomaly factor well known in quantum field theory.

\subsection{Hunt for the Red October?}
This is of course an interesting question. Clearly the $\hat F_i$'s are objects of the universe, it is perhaps one way to explain why our universe is a quantum universe.  In his later years, Heisenberg proposed the notion that Quantum Mechanics could fail as the energy involved in interactions increases. Dirac on the other hand conjectured that because the fine structure constant $\alpha=\frac{e^2}{\hbar c}$ is dimensionless approximately $\frac{1}{137}$, he voted that $\hbar$ can be got rid of in a better theory\cite{Dirac1963}. A convenient hypothesis may be the following:
\begin{equation}
{\hat F}({\hat p},{\hat q})=[{\hat p},{\hat q}]+i\hbar {\hat 1}+A {\hat p}+B{\hat p}^2-C{\hat p}^3-D{\hat p}^4-A' {\hat q}+B'{\hat q}^2+C'{\hat q}^3+D'{\hat q}^4 -...
\label{Hunting1}
\end{equation}
We eliminate cross-terms to avoid symmetrization problems \cite{Zoran2003}, but a casual look at this naive hypothesis of terms in the total Hamiltonian ${\hat H}_T$  already reveals some maybe interesting cosmological features. Is the $A$ constant which gives a background velocity related to the expansion of the universe? Is the B constant a renormalisation of the particle mass related to Mach's Principle? Is the $A$' constant (which gives a background acceleration) related to the universe's accelerated expansion like a cosmic Stark effect? Is the $B$' constant a cosmic harmonic force, while $C$' and $D$' are cosmic anharmonic forces and so on?  As noted earlier in section \ref{HeisenbergSummary}, Heisenberg was very concerned about anharmonic forces in his founding paper \cite{Heisenberg1925}.  From the empirical perspective, current technology can produce quantum dots with single electrons trapped in quantum wells. These are ideal candidates to search for the $\hat F_i$'s. Without doing more number crunching, we cannot tell how big a challenge this enterprise would be. With the fabrication of millions, perhaps even billions of quantum dots, subtle effects beyond the Heisenberg class such as depicted in eqn(\ref{Hunting1}) may be detectable?  This is an aspiration for the future. Interestingly in 1999 much controversy arose and still exists from the announcement of results of the Karl Popper experiment by Korean physicists Y-H. Kim and Y. Shih \cite{KIMShih1999} then at Cornell. Popper's experiment, first proposed in 1934 in his book Logik der Forschung (German edition), was a predecessor of EPR which like the latter does not involve spin, unlike most subsequent studies inspired by Bohm's book of 1954.  In 1935, after the now famous Einstein-Poldolsky-Rosen (EPR) paper was published, Einstein wrote a long letter to Popper criticising his experiment; this caused Popper to abandon his proposal.  Einstein, Poldolsky and Rosen did not cite Popper's experiment in their celebrated EPR paper.  As Poldolsky was responsible for most of its writing this may have been an oversight \cite{Pais1982}.  Einstein and Poldosky in fact fell out after the EPR paper. Einstein's hand written letter and its translation can be found in a report of an interview Popper gave in Paris just a few years before his death to Marie-Christine Combourieu in 1992 \cite{Mzarie-Christine1992}. In this letter, according to Popper, there was no definition of ``reality" as in the EPR paper. Sadly Popper died before the announcement of the results of the first realisation of his experiment, made possible only by advancement in parametric down-conversion and co-incident photon counting techniques down to single photons \cite{KIMShih1999}. The surprising result, still disputed, is that it seemed to agree with Popper's prediction that was designed to refute the Copenhagen interpretation of Quantum Mechanics. Without becoming involved in the controversy, it would be interesting to research if the Y-H. Kim and Y. Shih \cite{KIMShih1999} experiment or a variant of it could be used to detect features beyond Heisenberg's quantum mechanics.  In any case refinement of these experiments will surely take place over the next 100 years.

\section{D'Alembert and Einstein's Quantum Riddle}
\label{DAlembert-Einstein}
 Let's now return to d'Alembert's principle eqn(\ref{dAlembert2}). In classical mechanics ``harmony with given kinematic constraints'' is generally built into the equation by the use of generalised co-ordinates appropriate to the constraint, see remarks before eqn(\ref{dAlembert1}). However in the expression for our quantum d'Alembert's principle (see eqn(\ref{dAlembert3})) below we can introduce a quantum force ${\bf \hat \Phi}_n$ into eqn(\ref{dAlembert2}) as an additional inertia force:
\begin{equation}
\sum_{n=1}^{N}({\bf \hat F}_n-m_n {\bf \hat A}_n-{\bf \hat \Phi}_n)\cdot \delta{\bf \hat R}_n=0  \ .
\label{dAlembert3}
\end{equation}
The recovery of the force ${\bf \hat \Phi}_n$ from ${\hat F}_n$ is by no means trivial even if we know the latter such as our hypothesis eqn(\ref{Hunting1}), considering the complicated algebra generated by eqn(\ref{Hunting1}).  On the contrary, {\it if}\ we hypothesize the force we can reconstruct the function, subject to certain assumptions such as that the force ${\bf \hat \Phi}_n$ is monogenic and is related to the function ${\hat F}_n$ in the usual way \cite{Lanczos1986}. In 1946 the Hungarian physicist Imre F\'enyes seemed to have been the first to deduce the Schr\"odinger equation from classical stochastic mechanics \cite{Fenyes1946,Fenyes1948}.  This was further taken up in 1966 by the mathematician Edward Nelson \cite{Nelson1966} who provided a mapping of quantum mechanics to a classical stochastic model, albeit for Hamiltonians with quadratic momentum terms only. He obtained {\it both} the time-independent and time-dependent Schr\"odinger equations, effectively removing all the operator hats in eqn(\ref{dAlembert3}) with the force ${\bf \hat \Phi}_n$ postulated as a random Langevin force which defines a Markov process. The equivalence is striking and has since motivated a lot of research into stochastic models of quantum mechanics in spite of various limitations. Mapping from classical stochastic models to matrix algebra, already well known for Markov processes, see for example van Kampen \cite{vanKampen1980}, is relevant to quantum mechanics.  In fact this has been constructed for the F\'enyes-Nelson stochastic dynamics model by Davidson in 1979 \cite{Davidson1979}.  Unfortunately none of the above authors started from the d'Alembert's principle which is  deeper and in the quantum form eqn(\ref{dAlembert3}) is a statement akin to a {\it detailed balance} principle.  In particular in the classical case, the constraint force ${\bf \hat \Phi}_n$ must do no work for all virtual displacements \cite{Lanczos1986} and this must be strictly maintained upon quantization in the operator form. Moreover in accordance with Ehrenfest's theorem, the average $<{\bf \hat \Phi}_n\cdot \delta{\bf \hat R}_n>=0$, which implies that this force can only generate states in the Hilbert space that must be mutually orthogonal.  Understanding this will throw light on the origin of quantum fluctuations and non-locality, our aspired solution fot the full class I problem.  It is unclear at this stage what else can be learnt from more sophisticated classical stochastic models about the Heisenberg Group $\hat \chi$'s let alone going beyond in the pursuit to mimic quantum mechanics. Nevertheless these studies are useful illustrations for the interpretation of quantum mechanics and the attack on the measurement problem, see for example Weinberg \cite{Weinberg2014}.  In the next 100 years, we may finally be able to understand quantum mechanics, contrary to Richard Feynman's claim.

\section{Conclusion}

{\it Raffiniert ist der Herr Gott, aber boshaft ist Er nicht.} translated as: Subtle is the Lord, but malicious he is not - Albert Einstein.

In this paper we have given an appraisal of Heisenberg's great discovery of 1925 which laid the foundation of quantum mechanics. We have looked at some logical gaps in the foundation left by the early pioneers, notably Heisenberg, Born, Jordan, Dirac and Wigner. We have provided a response to Born's challenge and Einstein's riddle as to the origin of the fundamental Born-Jordan-Heisenberg quantum rule. We have extended our arguments from bosons to fermions by the introduction of Lagrange bracket quantization, which was missed by Dirac and have finally examined what could lie beyond Heisenberg's quantum mechanics.  In this we are led to consider if the fundamental quantum constraint function $\hat F_j$ which subtly underlies the basic quantum rule eqn(\ref{BJ Rule}) is in fact a cosmological object. Indeed if this is so, it will have major implications for any application of quantum mechanics to cosmological studies, such as black holes and dark matter etc. Over the last hundred years there have been many attempts at going beyond Heisenberg's quantum mechanics.  The author is not competent to provide even a glancing review of the wide subject. However there are a few that should be mentioned.  In 1979, Tom Kibble suggested a program to `geometrize' quantum mechanics \cite{Kibble1979}. This builds upon earlier works by mathematicians such as Bertram Kostant and Jean-Marie Souriau in the 1970s which use a generalization of Hilbert space and a symplectic structure as in classical mechanics. An excellent textbook on geometrical quantization which appeared in the last decade of the 20th century is given by Woodhouse \cite{Woodhouse1991}.  It is an active field of mathematical research which aims to extend quantum mechanics to treat non-linear relativistic systems, and `` rewrite quantum mechanics in a form better suited to unificaton with general relativity" \cite{Kibble1979}, an objective that remains elusive.  The same is true of other ``deformation" theories of geometrical quantization which date back to the early $\star$ product phase space formulation of quantum mechanics by Groeneweld \cite{Groenewold1946} and Moyal \cite{Moyal1949}. These techniques have nevertheless been embraced in recent years by quantum chemists because of their classical-like phase space approach and computational convenience, see for example Zachos et al \cite{Zachos 2005}. In recent decades too there have also been sophisticated generalizations of the Markovian master equation approach due in particular to G\"oran Lindblad \cite{Lindblad2020}, see for example Breuer and Petruccione \cite{Breuer2003}. Though originally developed to deal with open systems with non-unitary evolution due to the environment, it has now also attracted particle physicists to the field, such as Weinberg \cite{Weinberg2014}. I do not know if any of these studies are related to the ideas presented here but certainly there is much work for future generations to consider.

\section{Personal Tribute and dedications}
This work is a tribute to my many teachers, friends and colleagues, most of whom have since departed to paradise. Over the years, I have had the good fortune to learn from them, and to share with them my own journey towards an understanding of quantum mechanics. Alfred Lande taught me to question the foundations of quantum mechanics at a very young age at 21. John Valatin taught me all I needed to know about superconductivity and that it does not matter if a problem has been solved, the only thing that does matter is {\it how much} of it has been solved. Derek Martin taught me all I had to know about classical and quantum solid state physics. John Charap  taught me all about relativistic quantum mechanics. As he was a student of Paul Dirac then I must have got most of it all from the horse's mouth. Rudolph Peierls, our neighbour down the street at Oxford during my Harwell days, taught me a lot about physics, including his work with Landau on the measurability of quantum fields which angered Bohr. Wim Caspers shared with me the great joy of finding exact solutions to quantum many body problems, one of my favourite was the exact solution of the Thirring model.  I still remember how stunned I was when Walter Thirring gave a seminar in our department and we talked, I was just a student then.  Geoffrey Sewell taught me a lot about C* algebra and quantum statistical  mechanics but failed to convince me to become a mathematician.   Cyril Domb was like a father to many of us in our early graduate student days. Nicolaas van Kampen introduced me to the subtleties of Lagrangian constraints in classical and quantum mechanics at my first Dutch summer school in 1980 at Enschede. Tom Kibble taught me a lot about quantum field theory, he was a true gentleman and an inspiration while I was at Imperial College.  Moses Blackman, another student of Max Born's and Heisenberg was his PhD examiner, had his office next to mine at Imperial College. It was on a Sunday afternoon in 1983 when he met me in my office and announced that he had just finished his article on the history and physics of the magnetism of lodestones \cite{Blackman1983}. He told me something about it and that I must read it. Later that week, he passed away at 91.  Many would know that it was Heisenberg and his quantum mechanics that provided the first solution to the millennia old puzzle for the origin of ferromagnetism. Walter Kohn a student of Julian Schwinger was a personal friend and mentor through one of my difficult times.  He taught me a lot about quantum condensed matter theory; our friendship and mutual correspondence lasted until his death in 2016. Marshall Stoneham took me into Harwell in 1988 and we started a long collaboration until his death in 2010. Peter Schofield, Alan Lidiard and Norman March were old friends and mentors, they gave me many fond memories at Harwell. It was there that I regularly met with Nevil Mott who said he enjoyed talking to me even though my accent was funny. Roger Elliott was a wonderful friend who took the trouble to care about me and wrote to me during my difficulty time. Sam Edwards a great friend who hosted me for high table at Caius College several times, we had such fun together on other occasions. The multi-talented Peter Landsberg influenced me greatly, not only because he was one of my undergraduate examiners who supported me firmly even though I did not know it then. His astounding career as a mathematician, semi-conductor physicist and pioneering worker on the foundations of quantum mechanics taught me how boring life would be without all three.  Sadly, some of my  colleagues have also passed on, some far too early. Among them was Paul Kirkman (1986),  recently Jill Bonner (2021) and  Henk Blote (2022).

\section{Acknowledgement}

My thanks to Prof.  K.K. Phua, NUS and managing director of World Scientific (Singapore) Ltd for his trust in me to deliver this project.

\section{Appendix 1- Proofs of Dirac's Quantum Condition}
Dirac's sufficiency proof of his quantum condition \cite{Dirac1931} was an ingenious argument based on four operators ${\hat u}_1$,${\hat v}_1$,${\hat u}_2$ and ${\hat v}_2$.
These are all functions of the two canonical variables ${\hat p}$,${\hat q}$. Although not stated by Dirac explicitly we can now see that since there are only two canonical
variables ${\hat p}$,${\hat q}$ and the ${\hat u}$,${\hat v}$'s must be related to them via canonical transformations, then the four variables ${\hat u}_1$,${\hat v}_1$,${\hat u}_2$ and ${\hat v}_2$ must satisfy certain consistency relations between them when we generalise from classical variables to quantum operators. To do so Dirac exploited the Leibniz's rule for Poisson brackets: $\{uv,w\}=u\{v,w\}+\{u,w\}v$ and $\{u,vw\}=v\{u,w\}+\{u,v\}w$. Dirac then argued that if we have four operators in a Poisson bracket: $\{ {\hat u}_1 {\hat u}_2 , {\hat v}_1 {\hat v}_2 \} $, we have effectively two ways to evaluate this bracket. Equating the results of these two evaluations, Dirac obtained essentially:
\begin{equation}
\ \{ {\hat u}_1 , {\hat v}_1 \}[ {\hat u}_2 , {\hat v}_2]=[ {\hat u}_1 , {\hat v}_1]\{ {\hat u}_2 , {\hat v}_2 \}.
\label{Dirac Relation 1}
\end{equation}
If this relation must hold for ${\hat u}_1$,${\hat v}_1$ independent of ${\hat u}_2$,${\hat v}_2$, it is sufficient that:
\begin{equation}
\ \{ {\hat u}_1 , {\hat v}_1 \}=\kappa^{-1} [ {\hat u}_1 , {\hat v}_1] \ {\rm and} \{ {\hat u}_2 , {\hat v}_2 \}=\kappa^{-1} [ {\hat u}_2 , {\hat v}_2] ,
\label{Dirac Relation 2}
\end{equation}
with $\kappa=i\hbar$, which is Dirac's famous Quantum Condition \cite{Dirac1931}.  We shall now provide an alternative proof of this famous condition using only three operators instead of four. In this exercise we shall start with the operator Jacobi identity which is easily proved for any non-commutating set of operators:
\begin{equation}
\ [ {\hat a} , [ {\hat b},{\hat c}]  ]+ [ {\hat c} , [ {\hat a},{\hat b}]]+ [ {\hat b} , [ {\hat c},{\hat a}]] =0.
\label{Dirac Relation 3}
\end{equation}
Since we have already a one way proof of the Dirac quantum condition eqn(\ref{QuantumBrakets4}), we now have two ways to replace some of the commutators in eqn(\ref{Dirac Relation 3}) namely:
\begin{equation}
\ \{ {\hat a} , [ {\hat b},{\hat c}]  \}+ \{ {\hat c} , [ {\hat a},{\hat b}]\}+ \{ {\hat b} , [ {\hat c},{\hat a}]\} =0,
\label{Dirac Relation 4}
\end{equation}
or alternatively:
\begin{equation}
\ [ {\hat a} , \{ {\hat b},{\hat c}\}  ]+ [ {\hat c} , \{ {\hat a},{\hat b}\}]+ [ {\hat b} , \{ {\hat c},{\hat a}\}] =0,
\label{Dirac Relation 5}
\end{equation}
and they must be consistent.
Now subtracting equations (\ref{Dirac Relation 4}) and (\ref{Dirac Relation 5}), we now require that:
\begin{equation}
\ \{ {\hat a} , [ {\hat b},{\hat c}]  \}=[ {\hat a} , \{ {\hat b},{\hat c}\}  ] ,
\label{Dirac Relation 6}
\end{equation}
etc, for each of the remaining two cyclic permutations. We can now see that $\{\  ,\ \}=\kappa^{-1}[\ ,\ ]$  is a sufficient condition for eqn(\ref{Dirac Relation 6}) to be satisfied for by applying it we have $\kappa^{-1}[ {\hat a} , [ {\hat b},{\hat c}]  ]$ on the LHS while on the RHS we have $[ {\hat a} , \kappa^{-1} [ {\hat b},{\hat c}]  ]$ proving equality. Alternatively a direct expansion of eqn(\ref{Dirac Relation 6}) now gives the equations:
\begin{equation}
\ \{ {\hat a} , {\hat b}{\hat c}-{\hat c}{\hat b}  \}=\{ {\hat a} , {\hat b}{\hat c} \}-\{ {\hat a} , {\hat c}{\hat b} \}= \{ {\hat a} , {\hat b}\} {\hat c} + {\hat b}\{ {\hat a} , {\hat c}\}- \{ {\hat a} , {\hat c}\}{\hat b}- {\hat c}\{ {\hat a} , {\hat b}\} ,
\label{Dirac Relation 7}
\end{equation}
using Leibniz rule for the LHS and:
\begin{equation}
\ [ {\hat a} , \{ {\hat b},{\hat c}\}  ]={\hat a}\{ {\hat b} , {\hat c} \}-\{ {\hat b} , {\hat c} \}{\hat a} ,
\label{Dirac Relation 8}
\end{equation}
for the RHS. We now have an equation written only entirely in terms of Poisson brackets:
\begin{equation}
\{ {\hat a} , {\hat b}\} {\hat c} + {\hat b}\{ {\hat a} , {\hat c}\}+ \{ {\hat b} , {\hat c}\} {\hat a}= {\hat a}\{ {\hat b} , {\hat c}\}+\{ {\hat a} , {\hat c}\} {\hat b}+{\hat c}\{ {\hat a} , {\hat b}\}.
\label{Dirac Relation 9}
\end{equation}
Again a sufficiency condition for this to hold is of course Dirac's famous Quantum Condition written in the form: $\{ {\hat a} , {\hat b}\}=\kappa^{-1}[{\hat a} , {\hat b}]$, for all the Poisson brackets as easily verified.

\section{Appendix 2- Proof of Lagrange Bracket Quantum Condition}
In a similar way to Appendix 1, we shall use brackets involving three operators.  We start with the following operator Jacobi identity in terms of commutators $[\ , ]$ and anti-commutators $[\ , ]_+$ which is easily proved for operators:
\begin{equation}
\ [ [{\hat a} , {\hat b}]_+,{\hat c}]  ]+[ [{\hat b} , {\hat c}]_+,{\hat a}]  ]+[ [{\hat c} , {\hat a}]_+,{\hat b}]  ] =0.
\label{Appendix2-1}
\end{equation}
Next we need a super or graded Jacobi identity from classical mechanics involving both Poisson brackets $ \{ \ ,\}$ and Lagrange brackets $(\ ,) $ \cite{Goldstein1980} which can also be proved for operators, subject to symmetrization conditions \cite{Lemos2000, Choy2020a}:
\begin{equation}
\ \{ ({\hat a} , {\hat b}),{\hat c}\} + \{ ({\hat b} , {\hat c}),{\hat a}\} + \{ ({\hat c} , {\hat a}),{\hat b}\}=0.
\label{Appendix2-2}
\end{equation}
From Appendix 1 we can replace the Poisson brackets with commutators using the Dirac Condition:
\begin{equation}
\ [ ({\hat a} , {\hat b}),{\hat c}] + [ ({\hat b} , {\hat c}),{\hat a}] + [ ({\hat c} , {\hat a}),{\hat b}]=0.
\label{Appendix2-3}
\end{equation}
Equations eqn(\ref{Appendix2-1}) and eqn(\ref{Appendix2-3}) are now strikingly similar.  We now multiply eqn(\ref{Appendix2-3}) by a dimensional constant $\bar\kappa $ to ensure that they are dimensionally equivalent. Then they will be identical if the following holds:
\begin{equation}
\ \bar\kappa [ ({\hat a} , {\hat b}),{\hat c}] = [ [{\hat a} , {\hat b}]_+,{\hat c}] ,
\label{Appendix2-4}
\end{equation}
for each of the cyclically permutated terms. A sufficiency condition is the equality $( {\hat a} , {\hat b})=\bar{\kappa}^{-1}[{\hat a} , {\hat b}]_+$ with $\bar{\kappa}$ an appropriate dimensional constant. This completes the proof.

\section{Appendix 3- Proof of Jordan-Wigner Quantum Condition for fermion fields}
The non-relativistic second quantized Schr\"odinger field is given by the Hamiltonian:
\begin{equation}
H=\int \Bigl[ ({\hat \psi}^\dag \frac{-\hbar^2}{2m}\nabla^2 {\hat \psi} )+ {\hat \psi}^\dag V {\hat \psi}\Bigr] d\tau .
\label{Appendix3-1}
\end{equation}
Upon integration by parts and setting boundary terms to zero, we then have:
\begin{equation}
H=\int \Bigl[ \frac{\hbar^2}{2m} ({\nabla \hat \psi}^\dag {\nabla \hat \psi} )+ {\hat \psi}^\dag V {\hat \psi}\Bigr]d\tau  .
\label{Appendix3-2}
\end{equation}
We can now identify $\pi = -i\hbar \nabla \psi$ as the canonical field momentum for the Schr\"odinger field . Then the Lagrange bracket quantisation rule eqn(\ref{QuantumBrakets6}) is now given by:
\begin{equation}
[\hat \psi,\hat  \pi]_+= i\hbar (\hat \psi, \hat \pi)= i \hbar \delta(x-x')  .
\label{Appendix3-3}
\end{equation}
Here we have a factor of $i$ (which amounts to a choice for an arbitrary global gauge as in the boson case) since $\psi$ is no longer a Hermitian operator field. Now from the Lagrangian field density $L$ it is well known that:
\begin{equation}
\hat \pi = -i\hbar \nabla \hat \psi= \frac{\partial L}{\partial\dot {\hat\psi}}= i\hbar \hat \psi^\dag,
\label{Appendix3-4}
\end{equation}
see for example Heisenberg \cite{Heisenberg1929} or Schiff \cite{Schiff1949}.  With this convention we arrive immediately at the Jordan-Wigner field quantisation rule \cite{Jordan-Wigner1928} for fermions in the usual form:
\begin{equation}
[\hat \psi, \hat \psi^\dag]_+= \delta(x-x')  .
\label{Appendix3-5}
\end{equation}

\section{Appendix 4- Proof that $F_j$'s do not form a group unless its ghosts are Abelian}
The proof follows that of Dirac \cite{Dirac1961}, who had only considered classical constraints and their Poisson brackets. Consider an evolution in $\hat G$ induced by a change in $\hat \lambda$. From eqn(\ref{ReadOctober1}) we have:
\begin{equation}
\delta \hat G=\hat \epsilon_i \{\hat G,\hat F_i\},
\label{Appendix4-1}
\end{equation}
where $\hat \epsilon_i=\delta t(\hat \lambda_i-\hat \lambda_i')$ and we have dropped the $\circ$ reminder as understood. We can consider a further evolution induced by new $\hat \gamma_j$'s, so that:
\begin{equation}
\delta \hat G'=\hat \epsilon_i \{\hat G,\hat F_i\}+ \hat \gamma_j \{\hat G+ \hat \epsilon_i \{\hat G,\hat F_i\} ,\hat F_j\}.
\label{Appendix4-2}
\end{equation}
Now we can perform these evolutions in reverse order:
\begin{equation}
\delta \hat G''=\hat \gamma_j \{\hat G,\hat F_j\}+ \hat \epsilon_i \{\hat G+ \hat \gamma_j \{\hat G,\hat F_j\} ,\hat F_i\}.
\label{Appendix4-3}
\end{equation}
Now taking the difference and making use of the Jacobi identity we shall arrive at eqn(\ref{RedOctober3a}) after some algebra.


\begin{thebibliography}{10}
\bibitem{note1} The $n$ here is a quantum number borrowed from the Bohr model of the atom which specifies a particular orbit allowed by the old quantum rule of Bohr and Sommerfeld: $\oint p dq=J=nh$.
\end{thebibliography}
\end{document}